\documentclass[babel]{article-hermes}
\usepackage[T1]{fontenc}
\usepackage{epsfig}
\newenvironment{oupentry}[0]%
{ \setlength{\baselineskip}{14pt}%
  \begin{center}%
  \begin{footnotesize}%
  \begin{tabular}{p{10cm}}}%
{ \end{tabular}%
  \end{footnotesize}%
  \end{center}%
  \setlength{\baselineskip}{20pt}}

\newcommand{\mytilde}{$\sim$}

\def\figurename{}%

\begin{document}

\def\figurename{}

\renewcommand{\thefigure}{Figure \arabic{figure}}
%\renewcommand{\thetable}{Tableau \arabic{table}}

%\renewcommand \thetable {Tableau \arabic{table}}
%      \def\figname{figure}%
%      \def\tablename{Tableau}%
%      \def\tabname{tableau}%
%      \def\programname{Programme}%
%      \def\progname{programme}%sentre elle
%      \def\HermesRefname{Bibliographie}%
%\maketitle % Exemple de 1ere page sans les valeurs
%de champs
%            % pour afficher les commandes à
%utiliser

\journal[]%
   {TAL}%
   {42}{3}{2001}{667}{691}
        %}{}% Exemple avec numéros de page

\title[Dictionnaires et désambiguïsation lexicale]
      {Exploitation de dictionnaires électroniques \\ pour la désambiguïsation sémantique lexicale}

%\subtitle{Mode d'emploi des fichiers de style \\
%article-hermes.cls et biblio-hermes.bst}
\author{{\rm Caroline Brun --- Bernard Jacquemin --- Frédérique Segond}}

\address{%
Xerox Research Centre Europe\\
6. chemin de Maupertuis\\
38240 Meylan \\
\{Caroline.Brun,Bernard.Jacquemin,Frederique.Segond\}@xrce.xerox.com}

\resume{Cet article présente un système de
désambiguïsation lexicale sémantique, conçu
initialement pour l'anglais et à présent adapté à
la désambiguïsation du français. La méthodologie
développée repose sur l'utilisation d'un
dictionnaire électronique comme un corpus
sémantiquement étiqueté afin d'en extraire une base
de règles de désambiguïsation sémantique. Ces
règles permettent d'associer à un mot son sens
compte tenu de son contexte. L'extraction et
l'application des règles sont décrites en détail
ainsi que l'évaluation du système. L'évaluation des
résultats obtenus pour le français nous conduit à
considérer un ensemble de perspectives concernant
les ressources lexicales qui seraient les mieux
adaptées à la tâche de désambiguïsation dans le
cadre de cette méthodologie.}

\abstract{This paper presents a lexical
disambiguation system, initially developed for
English and now adapted to French. This system
associates a word with its meaning in a given
context using electronic dictionaries as
semantically annotated corpora in order to extract
semantic disambiguation rules. We describe the rule
extraction and application process as well as the
evaluation of the system. The results for French
give us insight information on some possible
improvments of the nature and content of lexical
resources adapted for disambiguation in this
framework.}
\motscles{désambiguïsation sémantique lexicale,
dictionnaire électronique, ressources lexicales.}
\thispagestyle{plain}
\maketitlepage
\section{Introduction}
Dans notre société, l'information -- et plus
particulièrement l'information électronique --
prend chaque jour plus d'importance, et sa maîtrise
est devenue la clef d'une certaine compétition,
tant au niveau politique qu'économique ou
scientifique. Dès lors, il est capital de pouvoir
gérer les masses de données sans cesse plus
abondantes qui sont mises à notre disposition. Dans
cette perspective, diverses applications
informatiques se révèlent indispensables pour qui
veut prendre part à cet essor.

En effet, l'analyse approfondie du contenu des
documents, la catégorisation de ce contenu, la
recherche d'une information précise dans une base
de données, ou encore la traduction assistée sont
des outils indispensables pour avoir une
compréhension globale et exacte des éléments jugés
intéressants \cite{VERONIS98}. Or ces différents
outils partagent un même postulat~: il s'agit de
pouvoir distinguer le sens adéquat de chaque mot en
fonction du contexte dans lequel il apparaît. C'est
là la tâche assignée aux systèmes de
désambiguïsation sémantique lexicale.

Cette désignation du sens idoine des mots en
contexte est également utile pour des logiciels
plus discrets tels que les correcteurs
orthographiques, aussi bien ceux des traitements de
texte que ceux qui permettent aux systèmes de
dictée vocale d'approcher une orthographe correcte.

Un désambiguïsateur sémantique lexical est un
système qui permet de sélectionner dans une liste
généralement prédéfinie la signification que
possède un mot polysémique en fonction de son
contexte d'apparition. Pour ce faire, plusieurs
méthodes existent qui tantôt sont concurrentes,
tantôt s'allient pour parvenir au meilleur
résultat. Deux types d'approches coexistent : les
modèles statistiques se basent essentiellement sur
l'étude de corpus sémantiquement étiquetés pour
déterminer le sens d'un mot polysémique en fonction
de son environnement lexical~; les approches
centrées sur les bases de connaissances consistent
à exploiter des ressources lexico-sémantiques
telles que les dictionnaires, lexiques ou
thesaurus. Généralement, toutes utilisent
l'environnement lexical des mots polysémiques.

%%Comment déterminer automatiquement le sens des
%mots polysémiques d'un texte en  fonction de leur
%contexte d'apparition~? Entre les logiciels d'aide
%à la traduction et ceux qui proposent une
%compréhension du sens des documents, nombreuses
%sont les applications qui demandent à identifier la
%signification des mots qu'elles traitent pour
%obtenir le meilleur des résultats escomptés.
%%Différentes méthodes existent et ont été testées
%qui tentent de sélectionner le sens le plus
%probable des mots d'un texte. Les uns utilisent
%pour ce faire des méthodes statistiques axées sur
%l'étude de grands corpus, d'autres exploitent les
%ressources sémantiques disponibles, thesaurus,
%dictionnaires ou lexiques. Généralement, tous
%utilisent l'environnement lexical des mots
%polysémiques.
La tâche de la désambiguïsation sémantique requiert
donc énormément de ressources lexicales sous la
forme de corpus sémantiquement étiquetés, de
dictionnaires, d'ontologies ou de réseaux
sémantiques. Or la carence de ces ressources est
omniprésente pour toutes les langues, excepté
l'anglais qui dispose de thesaurus comme le Roget,
d'ontologies sous formes de réseaux (WordNet, Cyc),
de corpus étiquetés (Brown, Hector).

Au cours des dernières années, de louables efforts
ont été menés pour que d'autres langues rattrapent
un peu du retard qu'elles ont pris dans la
construction de ressources lexico-sémantiques,
notamment avec l'initiative de RomansEval\footnote{
http://www.lpl.univ-aix.fr/projects/romanseval/}
qui a permis la constitution de corpus
sémantiquement étiquetés pour le français et
l'italien, ainsi qu'avec le consortium
EuroWordNet\footnote{http://www.hum.uva.nl/~ewn/},
chargé de construire un réseau sémantique
multilingue correspondant à son prédécesseur
anglais WordNet, et cela pour sept autres langues
(français, néerlandais, espagnol, italien,
allemand, tchèque et estonien). Ces initiatives
restent cependant ponctuelles et appellent des
mises en {\oe}uvre  plus importantes
(\cite{VIEGAS99}).

Par contre, les dictionnaires électroniques se sont
multipliés ces derniers temps, avec l'émergence des
langages à structure balisante qui permettent
d'organiser rationnellement l'information contenue.
Ainsi, les dictionnaires, disponibles dans de
nombreuses langues dans des formats SGML ou XML,
autorisent une exploitation précise des données
qu'ils contiennent et du fait de ces formats,
présentent une cohérence accrue par rapport à leurs
prédécesseurs  papier.

Cet article décrit un système de désambiguïsation
lexicale sémantique qui repose sur l'utilisation
d'un dictionnaire électronique et s'inscrit dans la
continuïté de précédents travaux dans ce
domaine\footnote{\cite{Segond2000} décrit une
exploitation des collocations et schémas de
sous-catégorisation d'un dictionnaire,
\cite{CBFS2000} présente un système équivalent à
celui décrit dans cet article et son évaluation
pour l'anglais, \cite{BRUN2000} détaille
l'architecture logicielle de ce système.}. Nous
avons en effet décidé de tirer parti de la
structure précise et de la cohérence accrue de ce
nouveau type de ressource. Notre système associe à
un mot un des sens que propose le dictionnaire en
fonction de son contexte lexico-syntaxique dans son
document d'apparition.  
Nous exposons tout d'abord des considérations sur
l'utilisation de dictionnaires dans le contexte de
la désambiguïsation lexicale sémantique et
décrivons en détail la ressource que nous avons
décidé d'exploiter.

Nous expliquons ensuite le fonctionnement du
système, c'est-à-dire les processus d'extraction et
d'application des règles.

Enfin, et c'est ce sur quoi nous souhaitons
insister dans cet article, l'évaluation du système
conduite pour le français nous amène à réfléchir
sur le type de ressource lexicale qui serait
appropriée pour mener à bien la tâche de
désambiguïsation.
\section{Dictionnaires}
\subsection{Dictionnaires et désambiguïsation
lexicale sémantique}

Comme nous l'avons indiqué dans l'introduction,
notre
système de désambiguïsation sémantique repose sur
l'exploitation d'un dictionnaire électronique. Les
dictionnaires constituent des ressources
particulièrement
intéressantes dans ce cadre. En effet,
\cite{Lesk86}, 
\cite{WilksAl90} et \cite{WILKS98} associent à
chaque
acception d'un mot le vocabulaire de sa définition
dans un
dictionnaire et exploitent ce vocabulaire pour
évaluer le
contexte dans un document et lui attribuer le sens
correspondant. \cite{VeronisIde90} enrichissent
cette
méthode en créant un réseau neuronal où chaque mot
est relié
à ses différents sens et où chaque sens est relié
aux mots
de sa définition, à leur tour reliés à leurs
différentes
acceptions. Certains (notamment \cite{GuthrieAl91}
et
\cite{CowieAl92}) développent cette même approche
en
s'écartant de la contrainte lexicale~: ils tirent
parti
d'une information supplémentaire de la version
électronique
du {\itshape Longman Dictionary of contemporary
English
\cite{LDOCE78}}, en particulier les primitives
sémantiques
classifiant les sens du lexique, et les
restrictions portant
sur les noms, les adjectifs et les arguments des
verbes en
fonction de leurs acceptions. Plus récemment,
\cite{Ellman00} et \cite{Litkowski00} ont entrepris
d'utiliser l'information lexicale, syntaxique et
sémantique
du dictionnaire Hector de SensEval pour effectuer
un
filtrage de sens par élimination des acceptions
inappropriées au contexte. Enfin, divers systèmes
utilisent
l'architecture hiérarchique de WordNet,
principalement les
différents sens définis ainsi que les relations
sémantiques
établies entre les n{\oe}uds, pour choisir celui
des sens
qui, en fonction de la structure de WordNet, se
rapproche le
plus de son contexte
(\cite{Sussna93,Resnik95,CarrollAl00,HawkinsAl00}).

%Comme nous l'avons indiqué dans l'introduction,
Notre système de désambiguïsation sémantique repose
sur l'exploitation d'un dictionnaire électronique.
Les dictionnaires constituent des ressources
particulièrement intéressantes dans ce cadre. En
effet, \cite{Lesk86},  \cite{WilksAl90} et
\cite{WILKS98} utilisent les définitions d'un
dictionnaire pour évaluer le contexte d'un mot et
lui attribuer un sens, tandis que
\cite{VeronisIde90} enrichissent cette méthode en y
adjoignant un réseau neuronal. D'autres exploitent
une autre information, notamment les primitives de
la version électronique du {\itshape Longman
Dictionary of contemporary English \cite{LDOCE78}}
(notamment \cite{GuthrieAl91} et \cite{CowieAl92}).
Plus récemment, \cite{Ellman00} et
\cite{Litkowski00} ont entrepris d'utiliser
l'information du dictionnaire Hector de SensEval
pour filtrer et réduire l'ensemble des sens
possibles des mots en contexte. Divers systèmes
exploitent également l'architecture hiérarchique de
WordNet pour définir une base de connaissances
(\cite{Sussna93,Resnik95,CarrollAl00,HawkinsAl00}).

Le choix d'une méthode basée sur l'utilisation d'un
dictionnaire comme ressource sémantique associée
repose sur une série d'arguments convaincants~: un
dictionnaire est conçu et organisé autour de la
notion de sens et de mot, et constitue de ce fait
un outil parfaitement adapté pour effectuer une
sélection du sens contextuel d'un mot. D'autre
part, le découpage des entrées polysémiques en
différents sens est extrêmement rigoureux, ce qui
implique une grande précision de cette information.
Dans cette perspective, on peut considérer le
dictionnaire comme un corpus sémantiquement
étiqueté particulièrement fiable.

Le dictionnaire que nous avons sélectionné, le {\it
Oxford-Hachette French
Dictionary} (OHFD, \cite{OHFD94}), contient de
très nombreux exemples représentatifs du sens qu'ils
illustrent, ainsi qu'une riche information sur les
collocations typiques d'un sens. D'autres informations comme les étiquettes
morpho-syntaxiques sont également très précieuses
dans le cadre de notre méthode. Le codage SGML dans
lequel ce dictionnaire a été conçu permet
d'appréhender aisément cette information pour la
traiter, et il permet, grâce à une définition de
type de document (DTD) stricte, d'identifier un
format de présentation de
l'information cohérent et aisément exploitable
(\cite{VERONISIDE98}).

De plus, ce dictionnaire est un ouvrage général,
qui reprend donc un maximum du vocabulaire courant
et de ses différentes significations. Il s'agit
d'un atout majeur pour un système de désambiguïsation
sémantique car il se rapproche ainsi d'une certaine
exhaustivité, tant au niveau des mots à
désambiguïser qu'au niveau des possibilités de
sélection. La méthode de constitution du
dictionnaire présente aussi un intérêt qu'il ne
faut pas négliger, puisque c'est à partir d'un
corpus que les sens ont été ordonnés dans chaque
entrée~: selon le corpus utilisé, une signification
qui se présente devant une autre a statistiquement
plus de chances d'apparaître que le sens suivant.
On verra dans la description de la méthode que ce
détail peut avoir son importance.

Enfin, c'est un dictionnaire bilingue, ce qui
présente évidemment des avantages pour les
applications de type multilingue, et notamment les
logiciels d'aide à la traduction. Mais un
dictionnaire bilingue contient certaines
particularités structurales intéressantes,
essentiellement un découpage en sens plus
systématique dans chaque langue du fait des
impératifs de correspondance avec l'autre langue.
Il s'ensuit une granularité plus fine, ainsi qu'un
nombre d'exemples significatifs plus important.

Bien entendu, la méthodologie proposée ici ne
nécessite pas impérativement l'utilisation d'un
dictionnaire bilingue. Elle peut aussi bien
s'appliquer en utilisant une ressource lexicale
monolingue ou même spécialisée si on veut se
concentrer sur un domaine précis. Toutefois, la
ressource retenue devra répondre à certaines
exigences~: il s'agit essentiellement de la qualité
des exemples et des collocations, qui devront être
représentatifs d'un sens auquel ils seront reliés
explicitement. En effet, c'est essentiellement de
ces informations que dépend la qualité des règles
de désambiguïsation, comme on le verra dans la
description du système.
\subsection{Description du dictionnaire OHFD}
Le dictionnaire OHFD comporte 47\,539 entrées pour
la
partie anglais-français et 38\,944 entrées pour la
partie français-anglais.
Le contenu du OHFD est découpé en cinq niveaux
imbriqués. Les vedettes correspondent aux entrées
du dictionnaire (1$^{er}$ niveau). Puis chaque entrée
est découpée en parties, selon les parties du
discours qui peuvent être celles de la vedette
(2\up{ème} niveau). 
Chaque partie du discours est ensuite découpée
selon les sens que peut
prendre la vedette (3\up{ème} niveau ou S1). Enfin
des indicateurs sémantiques
(4\up{ème} niveau ou S2) et/ou des indicateurs de
collocations et des exemples (5\up{ème} niveau ou
S3)
pointent sur des équivalents différents. 

Les éléments sont marqués  par leur fonction
(vedette, label, collocation, exemple, etc.).
Les étiquettes {\em sémantiques} du dictionnaire
qui permettent à
l'utilisateur de trouver l'équivalent qui convient
sont de trois
types\footnote{\/ Ces étiquettes sont décrites dans
\cite{AKROYD92}.} :
\begin{itemize}
\item les labels (balisés avec <la>) : ce sont des
marqueurs de
  registre/style (journalistique, familier, etc.),
de langue (anglais
  américain, écossais, etc.) ou de domaine
(architecture, médecine, zoologie, etc.);
\item les indicateurs (balisés avec <ic>) : ce sont
des notes explicatives
  qui précisent l'usage de l'entrée (par exemple,
pour l'entrée {\bf
    chien}, l'indicateur {\em de fusil} pointe sur
«~hammer~»).
\item les collocations (balisées avec <co>) : ce
sont les collocations habituelles 
  (sujet ou objet pour les verbes, nom modifié pour
les adjectifs).
\item les exemples : ce sont des exemples
d'utilisation de la vedette, qui illustrent le
contexte syntaxique et sémantique d'un sens donné
attaché à la vedette. Ces exemples peuvent être de
différents types :
\begin{itemize}
\item exemples de mot {\it composé} (balisés <lc>),
\item exemples {\it  idiomatiques} , (balisés <li>),
\item exemples {\it  phrasal verb} (pour
l'anglais), (balisés <lv>),
\item exemples de {\it structure}, (balisés <ls>),
\item exemples d'{\it usage}, (balisés <lu>),
\item exemples  {\it généraux}, (balisés <le>).
\end{itemize}
\end{itemize}

C'est l'information syntaxico-sémantique contenue
dans les indicateurs de collocations et les
exemples 
qui est exploitée par le système de
désambiguïsation que nous avons conçu. 

La figure \ref{tiede}  représente la version
«~papier~» de l'entrée lexicale du mot {\it
abandonner } dans le OHFD.

\begin{center}
\begin{figure}[p]
\begin{center}
\begin{oupentry}
abandonner /abA+~dOne/ 1 \\\\
I vtr \\\\

     1 ({\it renoncer à} ) to abandon, to give up
[{\it projet, théorie, activité, espoir}]; to give
up
     [{\it habitude}]; to give up, to forsake sout
[{\it confort, sécurité} ]; {\it Scol } to drop
[matière]; {\it \mytilde  les
     recherches } to give up the search;  {\it
\mytilde  la cigarette/l'alcool } to give up
     smoking/drinking;  {\it les médecins l'ont
abandonné} the doctors have given up on him;
     {\it je peignais, mais j'ai abandonné } I used
to paint, but I gave it up; {\it c'est trop dur,
     j'abandonne} it's too hard, I give up; {\it
\mytilde  la partie or lutte } to throw in the
towel ;\\
     2 ({\it céder} ) to give ou relinquish {\it
sout } [{\it bien} ] ({\it à qn } to sb); to hand
[sth] over
     [{\it gestion} ] ({\it à qn } to sb); {\it je
vous abandonne le soin d'expliquer} I'm leaving it
to you to
     explain; {\it elle lui abandonna sa main } she
let him take her hand ;\\
     3 ({\it se retirer de} ) to give up [{\it
fonction} ]; {\it Sport }({\it avant l'épreuve} )
to withdraw; ({\it pendant
     l'épreuve} ) to retire; {\it forcé d'\mytilde
la course} forced to withdraw from the race; {\it
\mytilde  ses
     études }  to give up one's studies ; \\
     4 ({\it quitter }) to leave [{\it personne,
lieu} ]; to abandon [{\it véhicule, objet, navire}
];{\it \mytilde  Paris pour Nice } to
     leave Paris for Nice; {\it il s'enfuit,
abandonnant son butin} he abandoned the loot and
     fled;{\it \mytilde  la ville pour la campagne
} to move out of town to live in the country; {\it
\mytilde  le
     terrain lit} to flee; {\it fig } to give up ;
\\
     5 ({\it délaisser} ) to abandon, to forsake
{\it sout } [{\it enfant, famille} ]; to abandon
[{\it animal} ]; to
     desert [{\it foyer, épouse, poste, cause,
parti}] ; \\
     6 ({\it livrer}) {\it \mytilde  qch à } to
leave ou abandon sth to; {\it \mytilde  un jardin
aux orties} to abandon a
     garden to the nettles; {\it  \mytilde  qn à
son sort } to leave ou abandon sb to his/her
     fate ; \\
     7 ({\it faire défaut} ) [{\it courage, chance}
] to desert [{\it personne} ]; {\it mes forces
m'abandonnent} my
     strength is failing me ; \\
     8 ({\it lâcher} ) to let go of [{\it outil,
rênes} ] ; \\
     9 {\it Ordinat} to abort .\\\\

II s'abandonner vpr \\\\

     1 ({\it se confier} ) to let oneself go ; \\
     2 ({\it se détendre} ) to let oneself go; {\it
s'\mytilde  dans les bras} de qn to sink into sb's
arms ;\\
     3 ({\it se laisser aller} ) {\it s'\mytilde  à
la passion/au désespoir} to give oneself up ou to
abandon\\
     oneself to passion/to despair; {\it s'\mytilde
au plaisir de } to lose oneself in the
     pleasure of; {\it s'\mytilde  au sommeil } to
let oneself drift off to sleep ; \\
     4 ({\it se donner sexuellement} ) [{\it femme}
] to give oneself ({\it à } to).
\end{oupentry}
\end{center}
\caption{Entrée de {\bf abandonner} dans le
Oxford-Hachette}
\label{tiede}
\end{figure}
\end{center}

\section{Description du système de désambiguïsation
lexicale sémantique}
\subsection{Architecture}
L'architecture du système de désambiguïsation
lexicale sémantique que nous avons 
développé est centrée sur l'utilisation du
dictionnaire OHFD utilisé comme un corpus 
sémantiquement étiqueté. Comme nous l'avons décrit
précédemment, pour chacun des mots du dictionnaire,
 les différents sens possibles sont distingués et
illustrés par des exemples d'utilisation, des
 indicateurs de collocations, des indicateurs
sémantiques etc. 
À partir de ces informations attachées à un sens
donné d'un mot, nous extrayons des règles de
désambiguïsation sémantique en utilisant un
«~shallow parser~», qui en extrait les relations
fonctionnelles ainsi qu'un réseau sémantique.  Ce
dernier permet d'associer des étiquettes
sémantiques aux arguments de relations
fonctionnelles. Ces règles prennent donc en compte
le contexte illustratif du sens du 
mot pour lequel elles sont construites. Une fois la
base de règles extraite pour l'ensemble du 
dictionnaire, elle est utilisable,  via un
algorithme spécifique d'application, pour la
désambiguïsation
des mots apparaissant dans du texte libre : les
phrases du texte sont analysées en utilisant le
même shallow parser et les relations fonctionnelles
en sont extraites. Pour chacun des mots entrant en
jeu dans une relation fonctionnelle on recherche
dans la base de règle si l'une d'entre elles peut
s'appliquer (la stratégie de sélection est
détaillée par la suite). Si c'est le cas, on peut
associer au mot le numéro de sens correspondant
dans le dictionnaire.
\clearpage
\begin{center}
\begin{figure}
\epsfxsize=11cm
\epsfysize=10cm
$$
\epsfbox{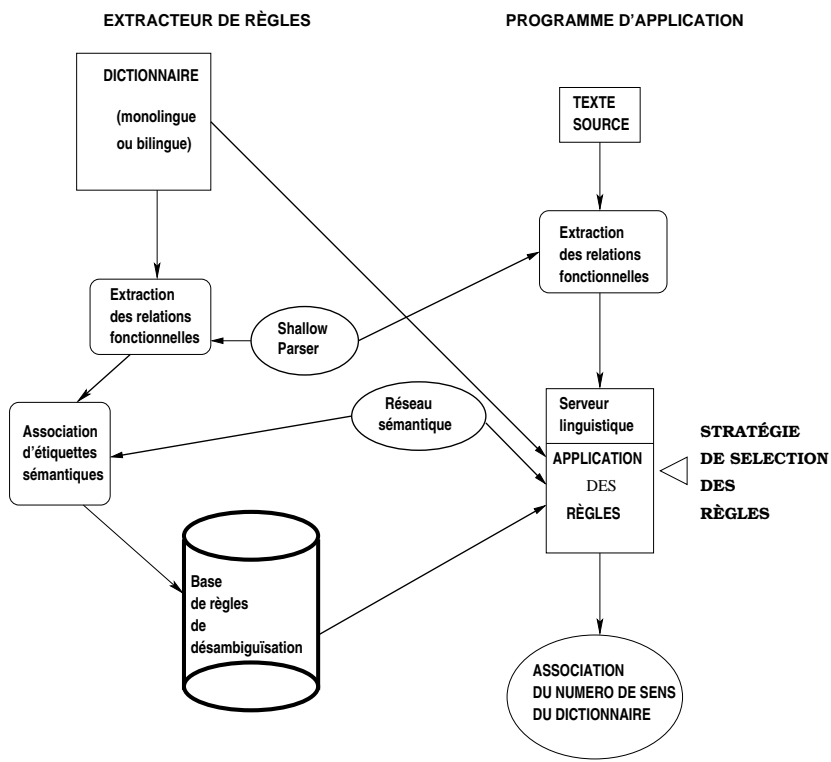} 
$$
\caption{Architecture du système}
\label{archi}
\end{figure}
\end{center}
L'architecture générale du système est illustrée
sur la figure \ref{archi}.
Ce système, initialement développé pour la
désambiguïsation sémantique de l'anglais
(\cite{CBFS2000,BRUN2000}), est maintenant adapté
au cas du français, ce que nous décrivons dans cet
article. 
\subsection{Shallow parser}
Les relations syntaxiques de type sujet, objet,
modifieur, etc., utilisées à la fois dans les phases
d'extractions et d'application des règles
sémantiques, sont extraites à l'aide d'un analyseur
syntaxique de surface («~shallow parser~»), IFSP,
\cite{Ait97}. La technologie de IFSP est fondée sur
des transducteurs d'états finis appliqués en
cascade pour marquer les syntagmes noyaux
(«~chunks~») et extraire les relations syntaxiques
entre syntagmes. Ces transducteurs sont appliqués
après une phase d'analyse présyntaxique :
segmentation, analyse morphologique et
désambiguïsation des parties du discours.

Voici un exemple d'analyse donnée par IFSP pour une
phrase tirée du journal {\it Le Monde}: \\ \\
{\it Un côté documentaire -- son réalisateur Marc
Levin y a fait ses premières armes -- et un côté
fiction qui nous fait penser qu'à chaque fois que
Slam s'aventure sur un terrain, il ne le fait qu'en
intrus, en abandonnant  les protagonistes à leur
sort, en oubliant plus ou moins consciemment des
règles élémentaires de mise en scène.} \\
{\small 
\begin{verbatim}
[NP Un côté NP]/N [AP documentaire AP] - [SC [NP
son réalisateur 
Marc_Levin NP]/SUBJ :v y a fait SC] [NP ses
premières armes NP]/OBJ
 - et [NP un côté NP]/OBJ [NP fiction NP]/N [SC [NP
qui NP]/SUBJ 
:v nous fait SC] [v penser v] [SC qu' [SC [PP à
chaque_fois_que Slam PP]
 :v s' aventure SC] [PP sur un terrain PP] , [NP il
NP]/SUBJ :v 
ne le fait SC] qu' en [AP intrus AP] , [v en
abandonnant v] 
[NP les protagonistes NP]/OBJ [PP à leur sort PP] ,
[v en oubliant v]
 plus_ou_moins consciemment [PP des règles PP] [AP
élémentaires AP] 
 [PP de mise_en_scène PP].

\end{verbatim}}
\hspace{-0.5cm}SUBJ(il,faire)  \hfill
SUBJ(Marc\_Levin,faire) \\
RELSUBJ(fiction,penser)  \hfill
RELSUBJ(fiction,faire) \\
DOBJ(abandonner,protagoniste)  \hfill
DOBJ(faire,côté) \\
DOBJ(faire,arme)  \hfill
VMODOBJ(oublier,de,mise\_en\_scène) \\
VMODOBJ(oublier,de=le,règle)  \hfill
VMODOBJ(abandonner,à,sort) \\
VMODOBJ(aventurer,sur,terrain)  \hfill
ADJ(premier,arme) \\
ADJ(réalisateur,Marc\_Levin)  \hfill
PADJ(règle,élémentaire) \\
PADJ(côté,documentaire)  \hfill NN(côté,fiction) \\
NNPREP(règle,de,mise\_en\_scène)  \hfill
NNPREP(protagoniste,à,sort) \\
\subsection{Extraction des règles de
désambiguïsation du dictionnaire}
Dans un premier temps, pour chacune des entrées du
dictionnaire, un numéro de sens (Snum) est associé
à chacune des différentes catégories sémantiques du
dictionnaire. Ces numéros de sens sont utilisés
comme des étiquettes sémantiques lors de
l'application des règles puisqu'ils désignent de
façon déterministe un sens spécifique d'une entrée
donnée. Dans le cas du OHFD, la numérotation
consiste à concaténer le numéro désignant la partie
syntaxique S1 (I, II, ...), et le numéro désignant
la partie sémantique S2 (1, 2, ...). 

Les règles extraites sont de deux types (cf. Ginger
I, \cite{DINI98,DINI99}):
\begin{itemize}
\item au niveau du mot (règles lexicales), ce sont
des règles qui s'appliquent sur la base du contexte
lexical du mot à désambiguïser;
\item au niveau des classes sémantiques (règles
sémantiques), ce sont des règles qui s'appliquent
sur la base du contexte sémantique du mot à
désambiguïser et qui sont construites en utilisant
les catégories sémantiques d'un dictionnaire ou
d'un thesaurus.
\end{itemize}

La base de règles est construite de la façon
suivante :
pour chaque sens Snum d'une entrée, les différents
exemples sont analysés syntaxiquement et les
relations en sont extraites au moyen de IFSP. Si
une relation met en jeu l'entrée elle-même, une
première règle de désambiguïsation est construite :
\\\\
{\bf Si l'entrée X apparaît dans la relation
syntaxique REL(X,Y), alors elle peut être
désambiguïsée avec son sens Snum.}\\\\Il s'agit d'une règle au niveau d'un mot
puisqu'elle considère le contexte lexical. 

Par exemple, dans l'entrée de {\it abandonner} (cf.
figure \ref{tiede}), un exemple de type général
illustre le sens I.6 de la section {\it vtr} : {\it
abandonner qn à son sort.}

Le shallow parser extrait les relations suivantes
de cet exemple : \\
DOBJ(abandonner,quelqu'un)\footnote{DOBJ =
complément d'objet direct.}  et
VMODOBJ(abandonner,à,sort)\footnote{VMODOBJ =
complément indirect modifieur du verbe.},
 ce qui nous permet de construire les règles
lexicales : \\\\
{\sf
abandonner$_{(I.1,I.2,I.3,I.4,I.5,I.6,I.7,I.8,I.9,II.1,II.2,II.3,II.4)}$
: }

\hfill {\sf DOBJ(abandonner,quelqu'un) => I.6 } \\
\\
{\sf
abandonner$_{(I.1,I.2,I.3,I.4,I.5,I.6,I.7,I.8,I.9,II.1,II.2,II.3,II.4)}$
:}

\hfill {\sf VMODOBJ(abandonner,à,sort) => I.6 } \\
\\
ce qu'on peut paraphraser ainsi : si le mot {\it
abandonner}, ambigu dans le OHFD entre les sens
(I.1,I.2,I.3,I.4,I.5,I.6,I.7,I.8,I.9,II.1,II.2,II.3,II.4),
apparaît dans les relations
DOBJ(abandonner,quelqu'un) et/ou
VMODOBJ(abandonner,à,sort), alors il peut être
désambiguïsé avec le sens I.6.

On construit des  règles semblables en utilisant
les collocations associées aux entrées lexicales.
Le type de la relation mise en jeu entre une
collocation et l'entrée lexicale qui lui correspond
est directement codée dans le dictionnaire. Par
exemple, dans une entrée pour un verbe, une
collocation est de type sujet ou objet, cette
information étant présente sous forme d'une balise
SGML.  On peut donc extraire les règles issues
d'information de type collocation directement à
partir des entrées, sans utiliser IFSP. 
 
Dans un deuxième temps, des classes sémantiques
sont utilisées afin d'élargir le champ
d'application des règles déjà construites. Pour
chacune de ces règles, l'argument Y de la relation
fonctionnelle est remplacé par l'ensemble des
classes sémantiques auxquelles il appartient :\\\\
{\bf Si l'entrée X apparaît dans la relation
syntaxique REL(X,classe\_d'ambiguïté(Y)), alors
elle peut être désambiguïsée avec son sens Snum.}\\

Ce sont des règles au niveau des classes,
puisqu'elles prennent en compte le contexte
sémantique du mot à désambiguïser.

Dans le cas du désambiguïsateur réalisé pour
l'anglais (\cite{CBFS2000,BRUN2000}), les classes
utilisées étaient les classes de plus haut niveau
données par WordNet. Pour l'adaptation du système
au français, s'est posé le problème du choix de
l'ontologie.
  Il n'existe actuellement aucune ressource
francophone comparable à  WordNet malgré
l'initiative du consortium EuroWordNet qui vise à
la  construction de réseaux sémantiques équivalents
à WordNet en diverses langues européennes, dont le
français. En effet, les objectifs de cette
entreprise ne visaient pas à décrire l'ensemble du
lexique, et le vocabulaire restreint du réseau ne
permet pas son expoitation dans le cadre d'une
application générale. Or, nous disposions d'un 
dictionnaire morpho-syntaxico-sémantique, le
dictionnaire AlethDic\footnote{Il s'agit de la
version 1.5.5 datant de 1994.} (\cite{Erli}), dont
nous pouvions exploiter l'information assez
directement.

Le contenu du dictionnaire électronique AlethDic
est une compilation de différents 
projets lexicographiques menés auparavant par
Gsi-Erli\footnote{Lexiquest à présent.}. La fusion
de 
plusieurs dictionnaires préexistants provenant
d'horizons différents a 
généré une grande variété de vocabulaire. 

AlethDic est une base lexicale (environ 55\,000
entrées) multicouche dont chacune des couches 
correspond à un niveau d'information linguistique~:
morphologique, 
syntaxique, sémantique. Ce type d'architecture
correspond aux 
recommandations GENELEX (\cite{Genelex}) édictées
dans le cadre du 
projet EUREKA (\cite{MenonAl93}) qui fait
correspondre à chaque entrée 
lexicale ses caractéristiques morphologiques,
qu'elles soient atomiques 
ou multiples. À chacune des unités morphologiques
correspond ensuite son 
ou ses pendants dans la couche syntaxique. Enfin,
chaque entité 
lexico-syntaxique est associée à une ou plusieurs
composantes 
sémantiques.

Dans le cadre de la désambiguïsation lexicale
sémantique, c'est la couche 
sémantique qui retient notre attention. Elle limite
ses informations aux 
noms, adjectifs et adverbes. Pour ces catégories,
l'information 
sémantique est de deux types~: le premier est la
description de chaque 
sens de l'unité lexicale à l'aide d'une classe
sémantique (3 catégories pour les adjectifs, 8 pour
les adverbes et 71 pour les noms) qui peut être
accompagnée d'un ou plusieurs traits~; le 
second correspond aux relations entre les unités
lexicales (antonymie, 
synonymie, etc.). C'est la répartition des entrées
en différentes 
catégories sémantiques que nous utilisons pour la
désambiguïsation. 

En utilisant cette ressource, les règles
sémantiques extraites sont les suivantes,
l'argument remplacé par la liste des classes
auxquelles il appartient étant respectivement {\it
quelqu'un} et {\it sort}:\\ \\
{\sf
abandonner$_{(I.1,I.2,I.3,I.4,I.5,I.6,I.7,I.8,I.9,II.1,II.2,II.3,II.4)}$
: }

\hfill {\sf DOBJ(abandonner,HUMAIN) => I.6 } \\ \\
{\sf
abandonner$_{(I.1,I.2,I.3,I.4,I.5,I.6,I.7,I.8,I.9,II.1,II.2,II.3,II.4)}$
:} 

\hfill {\sf
VMODOBJ(abandonner,à,EVENEMENT/ABSTRAIT) => I.6 }
\\ \\
ce qu'on peut paraphraser ainsi : si le mot {\it
abandonner}, ambigu dans le OHFD entre les sens
(I.1,I.2,I.3,I.4,I.5,I.6,I.7,I.8,I.9,II.1,II.2,II.3,II.4),
apparaît dans une relation DOBJ avec un mot
appartenant à la classe sémantique HUMAIN et/ou
dans une relation VMODOBJ avec un mot appartenant à
la classe sémantique EVENEMENT OU ABSTRAIT, alors
il peut être désambiguïsé avec le sens I.6. \\ 

Pour les deux sortes de règles,  le type de
l'exemple (<lc>,<le>,<li>,<lu>,<lv> ou <co> s'il
s'agit d'une collocation) est encodé dans la règle,
cette information étant utile lors de l'application
des règles (cf. infra). Chacun des exemples et
collocations attachés à un S2 donné possède
également un numéro, S3, que nous encodons
également.
 
Toutes les entrées du dictionnaire sont ainsi
traitées, ce qui permet la création d'une large
base de règles de désambiguïsation indexée sur
chaque mot (environ 217\,500 règles pour le
français et 270\,000 pour l'anglais\footnote{Cette
différence numérique est essentiellement due au
fait que nous ne disposons pas de classes
sémantiques pour les verbes.}). Cette base de
règles, extraite une fois pour toutes du
dictionnaire, est alors exploitable par le système
pour effectuer la tâche de désambiguïsation.
\subsection{Algorithme d'application}
La base de règles est  utilisée par un programme
d'application de règles qui permet la
désambiguïsation des mots d'un corpus quelconque.
Lorsque plusieurs règles sont en concurrence pour
un mot donné, une stratégie de sélection par type
de règle est appliquée. Dans le cas opposé où
aucune règle ne s'applique, le résultat donné par
défaut est le premier sens du OHFD (correspondant à
la partie du discours du mot). Comme ce
dictionnaire a été construit en utilisant les
fréquences des mots en corpus, le sens le plus
fréquent apparaît en premier. 

Tout d'abord la phrase contenant le mot à
désambiguïser est analysée syntaxiquement et les
relations syntaxiques en sont extraites. L'ensemble
des relations mettant en jeu le mot à désambiguïser
sont candidates pour être comparées aux règles de
désambiguïsation. 

On cherche en premier lieu les règles lexicales qui
peuvent s'appliquer, c'est-à-dire celles dont les
arguments sont exactement les mêmes, aux variations
morphologique près\footnote{Les arguments des
relations extraites par IFSP sont lemmatisés.}. 

Si une seule règle est applicable, elle est
sélectionnée et son numéro de sens est associé au
mot cible. 

Si aucune règle lexicale n'est directement
applicable, on cherche si l'on peut trouver des
équivalences entre règles et contexte syntaxique.
Par exemple, si une règle met en jeu une relation
de type sujet passif
SUBJPASS(A,B), elle est équivalente à une règle de
type objet direct DOBJ(B,A), et vice versa. Un
ensemble de relations syntaxiques sont
équivalentes, comme sujet SUBJ(A,B), sujet de
relative RELSUBJ(A,B) et agent passif PAGENT(B,A),
ou encore adjectif ADJ(A,B) et attribut ATTR(B,A).

Au contraire, si plusieurs règles lexicales peuvent
s'appliquer en concurrence, la sélection se fait en
se fondant sur le type de la règle, c'est-à-dire le
type de l'information à partir de laquelle elle a
été créée. Ainsi, les règles sont pondérées dans
l'ordre d'application suivant :
\begin{itemize}
\item règle issue de {\it collocation} (<co>), 
\item règle issue d'exemple de type {\it composé}
(<lc>), 
\item règle issue d'exemple de type {\it
idiomatique} (<li>),
\item règle issue d'exemple de type {\it  phrasal
verb} (dans le cas de l'anglais) (<lv>),
\item règle issue d'exemple de type {\it usage}
(<lu>),
\item règle issue d'exemple de type {\it général}
(<le>).
\end{itemize}

Cette stratégie repose sur les distinctions
linguistiques faites par les lexicographes au
moment de la construction du dictionnaire et prend
en compte la «~typicité~» des informations
attachées aux distinctions sémantiques des mots :
le choix de la règle à appliquer est réalisé en les
ordonnant de la plus typique à la plus générale. 

Cherchons à désambiguïser le mot {\it abandonner }
dans la phrase suivante : \\ \\
{\it Un côté documentaire -- son réalisateur Marc
Levin y a fait ses premières armes -- et un côté
fiction qui nous fait penser qu'à chaque fois que
Slam s'aventure sur un terrain, il ne le fait qu'en
intrus, en abandonnant  les protagonistes à leur
sort, en oubliant plus ou moins consciemment des
règles élémentaires de mise en scène.} \\

Les relations syntaxiques extraites par le shallow
parser et mettant en jeu le mot {\it abandonner}
sont :\\\\
DOBJ(abandonner,protagoniste)\\
VMODOBJ(abandonner,à,sort) \\\\
Ici, (au moins) une règle lexicale s'applique
directement :\\\\
{\sf
abandonner$_{(I.1,I.2,I.3,I.4,I.5,I.6,I.7,I.8,I.9,II.1,II.2,II.3,II.4)}$
:}

\hfill {\sf VMODOBJ(abandonner,à,sort) => I.6 } \\
\\
Le mot {\it abandonner}  peut donc être
désambiguïsé avec le sens I.6 du OHFD, qui a le
sens général de {\it livrer}. Comme le système
dispose du numéro d'exemple (S3), il propose la
traduction correspondante, ici {\it to leave ou
abandon sb to his/her fate}.

 Remarquons que plusieurs règles peuvent désigner
le même numéro de sens : c'est le cas en
particulier lorsque l'on peut extraire plusieurs
règles d'un exemple du dictionnaire et que l'on
retrouve exactement le même contexte dans la phrase
à traiter. La stratégie d'application prend en
compte cette possibilité en additionnant les poids
de ces règles, ce qui équivaut à faire prévaloir  les
conjonctions logiques de règles. 
 
Si aucune règle lexicale n'est susceptible de
s'appliquer, le système explore alors la liste des
règles sémantiques.  Afin d'effectuer la sélection,
la distance entre la liste L1 des classes
sémantiques d'une règle potentielle et la liste L2
des classes sémantiques associées à l'argument de
la relation en contexte est calculée comme suit : 

\begin{center}
$d=\frac{(CARD(UNION(L1,L2)) -
CARD(INTER(L1,L2)))}{CARD(UNION(L1,L2))}$
 
\end{center}
\vspace{0.5cm}

Cette distance peut varier de 0 a 1, 0 lorsque la
totalité des classes est commune, 1 lorsque
l'intersection est vide : la règle choisie est
celle qui présente une distance minimale. Dans le
cas où plusieurs règles présentent cette distance
minimale, la stratégie de sélection en fonction du
type est à nouveau appliquée. 

Cherchons à nouveau à désambiguïser {\it
abandonner} dans l'exemple suivant : \\ \\
{\it Sans tribune et sans reconnaissance légale, M.
Gustavo Arcos, secrétaire général du Comité pour
les droits de l'homme, s'efforce, sous la seule
protection des ambassades, de convaincre les jeunes
de ne pas abandonner le pays.}\\

Une relation syntaxique mettant en jeu le mot {\it
abandonner} est extraite par le shallow parser
:\\\\
DOBJ(abandonner,pays) \\

Dans le cas de cet exemple, il n'y a pas de règle
lexicale qui puisse s'appliquer. Par contre, les
classes sémantiques de {\it pays} étant
ESPACE\_LOCATIF, GEO, HUMAIN\_COLLECTIF, une règle
sémantique, extraite du sens I.4 de {\it
abandonner} dans le OHFD, est applicable. Cette
règle correspond à la collocation objet {\it lieu}
de {\it abandonner}, les classes sémantiques
attachées à {\it lieu}
étant : ENTITE, ESPACE\_LOCATIF, ANIMAL :\\\\
{\sf
abandonner$_{(I.1,I.2,I.3,I.4,I.5,I.6,I.7,I.8,I.9,II.1,II.2,II.3,II.4)}$
:}

\hfill {\sf
DOBJ(abandonner,ENTITE/ESPACE\_LOCATIF/ANIMAL) =>
I.4 }\\ 

Cette règle s'applique avec une distance de 0.8 par
rapport au contexte.
Le mot {\it abandonner}  peut donc être
désambiguïsé avec le sens I.4 du OHFD, dont le sens
général est {\it quitter} et la traduction {\it to
leave}\footnote{Dans le cas de règles sémantiques,
il nous a paru préférable de donner les sens et
traduction générales.}. 

Il peut arriver que plusieurs règles
potentiellement applicables désignent le même
numéro de sens. La stratégie de sélection prend en
compte cette possibilité en multipliant entre elles
les distances de ces règles ce qui a pour effet de
produire une distance «~globale~» diminuée, et donc
une plus grande probabilité de sélection. Le
système met donc l'accent sur la redondance
sémantique exhibée par le contexte du mot
(conjonction de règles).

\subsection{Implantation}
Le système présenté est implanté dans le cadre
d'une application client/serveur dédiée à l'analyse
linguistique. L'extracteur de règles est un client
qui utilise les différentes fonctionnalités du
serveur (principalement analyse syntaxique par le
shallow parser et consultation du dictionnaire). Le
programme d'application des règles est implanté
comme un service spécifique de consultation du
dictionnaire : la consultation sémantique du
dictionnaire. Lorsqu'un mot est désambiguïsé,
l'entrée correspondante est réordonnée en fonction
du sens qui lui est associé, ce sens étant présenté
en premier à l'utilisateur. Si toutefois la méthode
ne s'applique pas, le premier sens proposé est le
sens par défaut donné par le OHFD. Le système, qui
traite tous les mots d'un corpus quelconque, est
robuste et capable d'analyser  de larges corpus. 
\section{Évaluation des résultats}
\subsection{Évaluation}
Le système présenté dans cet article a précédemment
été évalué pour l'anglais \cite{BRUN2000}, dans le
corpus de SensEval \cite{Kilgarriff99}. Nous
souhaitons présenter ici les résultats obtenus
récemment pour la désambiguïsation du
français\footnote{Un système préexistant pour le
français a déjà été évalué dans \cite{LUX99}.}.
L'évaluation des résultats ne fait pas la
distinction entre
les erreurs propres au système de désambiguïsation
sémantique et les outils qu'il utilise dans sa
chaîne
d'analyse. Nous désirions en effet connaître les
résultats
réels de notre système plutôt que le résultat
théorique lié
exclusivement à la désambiguïsation lexicale
sémantique.

Le but de cette évaluation est double : d'une part
identifier les problèmes liés à la méthode et aux
ressources utilisées, d'autre part,  constituer un
corpus sémantiquement étiqueté avec les sens du
OHFD. Le découpage d'un mot en sens pose souvent
des problèmes que les lexicographes discutent entre
eux. Dans l'utilisation
que nous faisons d'un dictionnaire, nous avons opté
pour la confiance dans le choix des concepteurs des
ressources utilisées \cite{VeronisT98},
\cite{Segond2000b}. 
L'évaluation du système pour le français est
conduite en utilisant pour corpus un extrait du
journal {\it Le Monde}
d'environ 850 phrases\footnote{Il s'agit d'une
évaluation partielle, le corpus choisi comptant au
total environ 3\,000 phrases.}.
Les mots qui nous intéressent lors de cette
évaluation sont les noms, les adjectifs et les
verbes, ce qui représente un total d'environ
14\,000 mots. Nous obtenons les résultats de
précision et couverture présentés sur le tableau
\ref{tab1}. Ces résultats sont donnés en fonction
des parties du discours ainsi que du type de règle
(lexicale ou sémantique).

\begin{table}[ht]
\begin{center}   
\begin{tabular}{|*{7}{l|}}
\hline
Cat.  & Préc.    & Préc.          & {\bf Préc. }
& Couverture & Couverture & {\bf Couverture  }\\
     & Règles    & Règles        & {\bf toutes }
& Règles     & Règles     & {\bf toutes } \\
     & Lex.      & Sém.          & {\bf Règles }
& Lex.       & Sém.       & {\bf  Règles } \\
\hline
Noms & .88        & .48        & {\bf .68 }
& .24        & .14        & {\bf .38  }\\
\hline
Verbes & .97      & .50         & {\bf .58 }
& .08       & .19         & {\bf .27 } \\
\hline
Adj.   & 1        & .51         &  {\bf .68  }
& .23       & .23         & {\bf .46  }\\
\hline
{\bf Total } & {\bf .90  }      & {\bf .50  }
& {\bf .65  }         & {\bf .19  }      & {\bf.16
}        & {\bf.35 } \\
\hline
\end{tabular}
\caption{Résultats de l'évaluation pour le
français}
\label{tab1}
\end{center}
\end{table}

%\vspace{0.5cm}

Plusieurs remarques peuvent être faites concernant
ces résultats :
\begin{enumerate}
\item D'une manière générale, la précision globale
obtenue
par le système (.65) dans le cas du français est
nettement plus faible que celle obtenue pour
l'anglais (il s'agissait
d'environ .80). La précision fournie par les règles
lexicales étant tout à fait comparable (.90 dans
les deux cas), on constate donc que le problème
provient principalement des règles sémantiques dont
la précision est relativement médiocre pour le
français (.50 contre environ .70 pour l'anglais).

Il semble donc que l'utilisation des classes
données dans WordNet  est plus adaptée à la tâche
de désambiguïsation que celles proposées par le
dictionnaire AlethDic. 

En particulier, l'information lexico-sémantique
d'AlethDic est 
constituée de divers lexiques disparates
préexistants qui sont souvent 
spécifiques à un domaine d'application. Or, si le
domaine marque 
parfois explicitement la signification spécialisée
d'un mot, ce n'est 
pas une constante.

Le mot {\itshape condition} dispose des
classes\footnote{Cf. définitions libres AlethDic.}
 ENTITE (au sens de «~base d'un accord~»), ABSTRAIT
(au sens de «~éléments, circonstances qui 
déterminent une situation~»), mais aussi ESPACE (
«~Lieu où l'on pratique 
le conditionnement d'un textile~»).

Or, ce dernier sens est spécialisé et il ne
correspond à aucune 
signification du OHFD. Le manque de distinction
entre le vocabulaire 
général et spécialisé, ainsi que l'incohérence
entre AlethDic et OHFD 
sur ce sujet, est source d'erreur lors de
l'application des règles 
sémantiques.

D'autre part, on retrouve, comme dans le cas de
l'utilisation de WordNet, le problème de classes
trop 
générales qui élargissent excessivement la portée
des règles sémantiques. Par exemple, la classe 
ABSTRAIT, à laquelle appartiennent {\itshape style}
et {\itshape dieu}, 
recouvre l'«~ensemble des notions universellement
reconnues~». En 
contexte, on pourra donc obtenir des erreurs de
confusion dues à une 
généralisation excessive. Dans la phrase~:\\\\
{\itshape Ils nous demandaient : «~C'est le soleil
ou c'est le bon Dieu ?~»}\\\\
le mot {\itshape bon} sera fautivement désambiguïsé
par la règle 
~:\\\\
{\sf
bon$_{(I.1,I.2,I.3,I.4,I.5,I.6,I.7,I.8,I.9,I.10,I.11,I.12,II.1,III.1,III.
2,III.3,IV.1,V.1)}$ }

{\hfill {\sf ADJ(bon,ABSTRAIT) => II.2}}\\\\
qui est la règle sémantique correspondant à la
règle lexicale ~:\\\\
{\sf
bon$_{(I.1,I.2,I.3,I.4,I.5,I.6,I.7,I.8,I.9,I.10,I.11,I.12,II.1,III.1,III.
2,III.3,IV.1,V.1)}$ }

\hfill {\sf ADJ(bon,style) => II.2}\\\\
Or, le sens désigné par cette règle de
désambiguïsation correspond au 
sens «~de qualité~» alors que «~bienveillant~»
aurait été plus indiqué. 
Cette erreur vient du fait que selon AlethDic,
{\itshape dieu} et 
{\itshape style} sont sémantiquement équivalents de
par leur classe 
ABSTRAIT. 

Il ne faudrait pas en conclure qu'AlethDic soit
meilleur que WordNet, ou l'inverse : c'est la
cohérence entre ressources dictionnairique et
thésaurique qui importe (cf. infra). Ce problème
de cohérence est d'ailleurs probablement mis en
exergue par la granularité plus fine des classes
sémantiques utilisées dans le cas du français.
\item La couverture globale (.35) est à peu près
semblable à celle que nous obtenions dans le cas de
l'anglais (qui était de .37), ce qui est
relativement étonnant puisque, ne disposant que des
classes sémantiques pour les noms et quelques
adjectifs et adverbes en français, le nombre de
règles sémantiques extraites dans le cas du
français est bien inférieur à celui de l'anglais
(86\,400 règles sémantiques en français contre
132\,000 en anglais).  Il semble que la couverture
des règles lexicales en français soit nettement
supérieure à celle de l'anglais, en particulier
pour les noms, peut-être est-ce dû à la nature des
corpus d'évaluation\footnote{HECTOR (cf. SensEval)
pour l'anglais, {\it Le Monde} pour le français.}.
\item Dans les deux cas, la couverture est
relativement faible.
 Une première remarque que nous pouvons faire est
que l'extraction des règles est loin de couvrir la
totalité des mots du dictionnaire. Le tableau
\ref{tab2} récapitule quelques chiffres concernant
le dictionnaire et les règles extraites.

\begin{table*}[h]
\begin{center}
\begin{tabular}{|*{4}{l|}}
\hline 
  & Nombre total & Nombre d'entrées   & Nombre
moyen  \\
  & d'entrées en & pour lesquelles    & de règles
par \\
  & français     & au moins une règle & mots
\\
  &           &  est extraite      &
\\

\hline
français & 38 944       & 15 224             & 14.3
\\
\hline
anglais  & 47 539       & 16 091             & 16.8
\\
\hline
\end{tabular}
\caption{Quelques données numériques associées au
dictionnaire}
\label{tab2}
\end{center}
\end{table*}
%\vspace{0.5cm}

Pour bien des entrées lexicales, ni exemples ni
collocations ne sont attachées aux différents sens
des mots, et nous ne pouvons donc pas en extraire
de règles. Le dictionnaire OHFD est en effet
construit pour des utilisateurs humains et bien
souvent un indicateur ou une étiquette sémantique
est suffisant pour caractériser un sens donné. 
\end{enumerate}
\subsection{Perspectives}
La méthodologie employée a montré de bons résultats
pour l'anglais et des résultats un peu moins bons
mais néanmoins intéressants dans le cas du français
: nous sommes à présent tout à fait convaincus de
l'intérêt de l'utilisation de l'information d'un
dictionnaire ainsi que du contexte syntaxique dans
le cadre de la tâche de désambiguïsation. En outre,
cette méthode a l'avantage de ne pas nécessiter de
corpus d'apprentissage, contrairement aux méthodes
statistiques par exemple : le dictionnaire fait
office de corpus sémantiquement étiqueté tout en
n'étant pas cantonné à ce rôle puisqu'ayant bien
d'autres usages possibles.    

Cependant la  méthode présente un certain nombre de
problèmes liés à la cohérence entre  informations
du dictionnaire et informations sémantiques des
thesaurus. Elle présente aussi un problème en termes
de couverture puisque celle-ci est relativement
limitée. Nous avons plusieurs idées qui nous
semblent intéressantes pour essayer de pallier ces
différents problèmes. 

Partons d'un exemple qui montre le besoin
d'améliorer la cohérence entre ressource lexicale
et classes sémantiques. Afin de désambiguïser {\it
assumer} dans la phrase suivante :\\\\
{\it 
Il s'agit aussi d'{\bf assumer} une mission
pédagogique vis-à-vis de l'opinion, 
mission que je crois essentielle, aujourd'hui,
compte tenu de la complexité des problèmes.}\\\\
le système applique la règle suivante : \\\\
assumer$_{I.1,I.2}$
DOBJ(assumer,ABSTRAIT/ENTITE/ESPACE/ETAT) => I.2
\\\\
Cette règle est construite en utilisant la
collocation objet du sens I.2 de {\it assumer} à
savoir {\it assumer sa condition}. Elle pointe sur
le sens {\it accepter} de l'entrée alors que {\it
prendre en charge} (I.1) aurait été plus indiqué
dans ce contexte.

Les mots {\it condition} et {\it mission} possèdent
en commun les classes ENTITE (décrivant les objets
abstraits) et ESPACE (comprenant les lieux
aménagés), et de ce fait le système les a
rapprochés. D'une part, ENTITE est une classe
extrêmement générale. De l'autre, ESPACE recouvre
une acception spécialisée de {\it condition} (lieu
où l'on pratique le conditionnement d'un textile),
qui est absente du OHFD ; quant à l'acception de
{\it mission} en tant que lieu dans le OHFD, elle
est spécifiée par une étiquette {\it religieux}, en
association avec d'autres significations ({\it
charge, organisation}), membres de la classe
ENTITE. 
On retrouve aussi le même genre de problème dans le
cas de l'anglais, les recoupements entre le OHFD et
WordNet n'étant pas toujours cohérents.

Afin d'améliorer le système nous pensons qu'il
serait intéressant d'utiliser le dictionnaire
lui-même comme ressource sémantique. Voici les
trois entrées lexicales (raccourcies) mises en jeu
dans l'exemple précédent :\\\\\\\\
{\bf mission} : /misjO+~/ nf \\\\
     I.1 (tâche) mission, task; ...\\
     I.2 (fonction temporaire) mission, assignment;
... \\
     I.3 (groupe) mission, team; ...\\
     I.4  Mil (but) mission ;  \\
     I.5 Relig (charge, organisation, bâtiment)
mission; ... \\ \\ 
{\bf assumer} /asyme/ vtr \\\\
     I.1 (prendre en charge) to take
[responsabilité]; to hold [fonctions]; ...\\
     I.2 (accepter) to come to terms with
[condition, identité, passé]; \\ \\
\hspace{-0.5cm}{\bf
responsabilité}/KEspO+~sabilite/ nf \\ \\
I.1 (participation) gén responsibility;\\
I.2 (charge) responsibility; \\
I.3 (fait de devoir répondre de ses actions)
responsibility; \\
I.4 Assur liability; ~ civile personal liability.
\\

Si l'on pouvait associer  de manière cohérente une
classification sémantique à chacun des sens du
dictionnaire, nous pourrions disposer d'une base de
règles cohérente et également beaucoup plus
générale que celle dont nous disposons à l'heure
actuelle. En utilisant les indicateurs et les
étiquettes sémantiques (dont un traitement
spécifique est également réalisé dans le cadre du
projet DEFI \cite{Michiels2000}), nous pourrions
envisager de trouver des classes sémantiques
générales valides sur la totalité du dictionnaire.

Par exemple, si nous ajoutons des classes
sémantiques caractérisant les différents S2 aux
entrées précédentes, indiquées par <csem> ...
</csem>, nous obtenons les entrées «~augmentées~»
suivantes : \\ \\
%\clearpage
{\bf mission} : /misjO+~/ nf \\\\
     I.1 <csem>rôle</csem> (tâche) mission, task;
...\\
     I.2 <csem>fonction</csem> (fonction
temporaire) mission, assignment; ... \\
     I.3 <csem>groupe</csem> (groupe) mission,
team; ...\\
     I.4 <csem>militaire</csem> Mil (but) mission ;
\\
     I.5 <csem>religieux</csem> Relig (charge,
organisation, bâtiment) mission; ... \\ \\ 
{\bf assumer} /asyme/ vtr \\\\
     I.1 <csem>s'occuper</csem>(prendre en charge)
to take [responsabilité]; to hold [fonctions];
...\\
     I.2 <csem>accepter</csem> (accepter) to come
to terms with [condition, identité, passé]; \\ \\
{\bf responsabilité}/KEspO+~sabilite/ nf \\ \\
I.1 <csem>morale</csem>(participation) gén
responsibility;\\
I.2 <csem>rôle</csem>(charge) responsibility; \\
I.3 <csem>juridique</csem>(fait de devoir répondre
de ses actions) responsibility; \\
I.4 <csem>assurance</csem> Assur liability; ~
civile personal liability.\\

Nous pouvons alors créer les règles suivantes pour
le sens I.1 de {\it assumer}\\\\
assumer$_{I.1,I.2}$ DOBJ(assumer,responsabilité) =>
I.1\\
assumer$_{I.1,I.2}$
DOBJ(assumer,morale/rôle/juridique/assurance) =>
I.1 \\

Ces règles nous permettent de désambiguïser {\it
assumer}, et en particulier de retrouver le sens
I.1 dans le contexte de notre exemple, le
complément d'objet {\it mission} ayant {\it rôle}
pour classe
sémantique.

On peut aussi créer un nouveau type de règle plus
général, basé uniquement sur les classes
sémantiques attachées aux sens des entrées :\\\\
{\sf V$_{s'occuper,accepter}$
DOBJ(V,morale/rôle/juridique/assurance) =>
s'occuper }\\ 

Ce qui peut être paraphrasé ainsi : si un verbe,
ambigu entre les classes {\it s'occuper} et {\it
accepter},
 a pour complément d'objet direct un mot ambigu
entre les classes {\it morale, rôle, juridique}, et
{\it assurance},
alors il peut être désambiguïsé avec son sens {\it
s'occuper.}\\

Les étiquettes sémantiques nous paraissent être des
sources d'informations qu'il serait valable
d'exploiter afin de dériver une hiérarchie
sémantique du dictionnaire.  Par exemple, on trouve
dans le OHFD les indicateurs suivants : {\it
garçon, fille, homme laid, vieille femme, homme
préhistorique ...}, qui pourraient constituer les
feuilles d'une ontologie spécifiant la classe {\it
humain}. Cette hiérarchie serait certainement
bénéfique à l'amélioration et à la généralisation
de notre application. Nous sommes conscients de la
difficulté d'une telle tâche dont la méthodologie
reste à élaborer. La conception d'un jeu
d'étiquettes cohérent couvrant l'ensemble du
dictionnaire constitue un sujet de recherche à part
entière.  L'automatisation complète d'un tel
processus s'est d'ailleurs déjà révélée
problématique : l'extraction automatique de larges
bases de connaissances à partir de dictionnaires
(\cite{Chod85,Klavans90,WilksAl90,Ide93}) n'a
jamais complètement abouti. Il semble a priori
qu'une phase de validation, voire d'enrichissement
manuel soit nécessaire, mais cet effort nous semble
justifié compte tenu de la qualité de la ressource
lexicale que l'on obtiendrait et de son utilité
dans le cadre d'applications de TAL.
\section{Conclusion}
Cet article  présente un système de
désambiguïsation lexicale sémantique s'appuyant sur
l'utilisation d'un dictionnaire électronique. Le
système extrait une fois pour toutes une base de
règles de désambiguïsation du dictionnaire; ces
règles sont par la suite utilisées pour
désambiguïser des mots dans des textes de nature
quelconque. Ce système présente différents
avantages : l'utilisation du dictionnaire évite de
recourir à des corpus sémantiquement étiquetés,
très difficiles à obtenir; les informations
exploitées sont fiables dans la mesure où elles
sont validées par des lexicographes; le contexte
syntaxique s'avère également très utile pour la
désambiguïsation; la taille des données traitées
est transparente pour le système, dans la mesure où
le contexte utilisé pour la désambiguïsation ne
dépasse pas une phrase. 

Cependant, un certain nombre de problèmes sont mis
en évidence dans la phase d'évaluation. Si les
règles lexicales offrent une très grande précision,
celle-ci doit être améliorée pour les règles
sémantiques. Une amélioration de la couverture du
système serait également la bienvenue.

Plusieurs axes de recherche sont envisagés. Nous
pensons tout d'abord modifier le système afin d'y
intégrer un analyseur syntaxique  plus élaboré,
XIP\footnote{En cours d'implantation à XRCE.}, qui
permettrait d'encoder dans les relations
syntaxiques plus d'informations sur le contexte,
par exemple la nature du déterminant introduisant
un groupe nominal (par exemple {\it abandonner
[+Humain] à [+Possessif] [ADJ] sort }). 

Ensuite, les règles de désambiguïsation conservent
les classes d'ambiguïtés des arguments des
relations. Mais dans un exemple du dictionnaire
comme {\it abandonner un lieu}, l'argument, ici le
mot {\it lieu}, appartient à une classe donnée
(dans ce cas ESPACE\_LOCATIF selon AlethDic, et non
ENTITE ou ANIMAL).
Si nous parvenions à isoler la classe sémantique
des arguments dans le contexte donné par le
dictionnaire, la précision des règles serait très
largement accrue.
 
Enfin, l'amélioration du système passe par une plus
grande cohérence entre les ressources sémantiques
et le dictionnaire :  il nous semble très
intéressant d'utiliser le dictionnaire lui-même
comme ressource sémantique afin de construire une
base de connaissances sémantiques. Une telle
ressource aurait plusieurs avantages dans le cadre
de l'application visée : elle offrirait un meilleur
recouvrement des sens dans les règles et
permettrait un élargissement de la portée de
celles-ci. En effet, nous pourrions créer des
règles additionnelles désambiguïsant non plus des
mots mais des classes de mots en fonction du
contexte.

D'une manière générale, ce système a de nombreuses
applications : nous l'utilisons déjà dans un
système d'aide à la compréhension de textes écrits
dans des langues étrangères. L'intégrer dans des
processus d'indexation sémantique et plus
généralement dans tout type d'application visant à
l'extraction et  la compréhension de connaissances
contenues dans des documents électroniques nous
semble particulièrement prometteuse.
\acknowledgements{Nous tenons tout particulièrement
à remercier chaleureusement Agnès Sandor pour son
travail sur l'évaluation du système, ainsi que pour
sa relecture attentive de cet article.}
\nocite{*}
\bibliographystyle{biblio-hermes}

%\bibliography{BrunJacquemSegond}
\end{document}